# The contribution of machine learning to the prevention of burnout among healthcare workers in Morocco

**First and last name :** EDDAOU Mohammed

**Function / Title:**Teacher-researcher

**Structure :**Laboratory "Arithmetic, Scientific Computing and Applications (ACSA)", Team "Number Theory, Cryptography and Computer Systems (TNCSI)", Faculty of Sciences, Mohammed Premier University, Oujda, Morocco.

**E-Mail :** mohammed.eddaou@ump.ac.ma.

**Abstract :**

In recent years, and particularly during the Covid-19 pandemic, Morocco has experienced significant pressure from user demand, leading to a significant workload in public hospitals. This situation raises major questions regarding the occupational health of healthcare staff. While previous studies have focused on the role of AI in the safety and resilience of military personnel, no research has investigated its role in protecting healthcare personnel from psychosocial risks. This inadequacy leads us to formulate the following central question:What is the contribution of machine learning to the prevention of emotional exhaustion (burnout) among healthcare staff in Morocco?

This work is part of a modeling approach aimed at developing a predictive model of the risks of emotional exhaustion (burn-out), the parameters of which will be estimated using supervised learning.

From a scientific perspective, this work aims to contribute to the development of systems for preventing psychosocial risks affecting staff in healthcare establishments.

From a managerial perspective, this research aims to equip decision-makers in healthcare establishments so that they can anticipate psychosocial disorders linked to emotional exhaustion (burn-out) and implement appropriate preventive measures.

## 1.Introduction

Morocco, in recent years—particularly during the Covid-19 pandemic—has experienced high user demand, resulting in a significant workload in its public hospitals (Kapasa et al., 2021; Fattahi et al., 2023). This situation raises questions about the occupational health of healthcare staff.

While previous studies have primarily focused on the role of AI in military safety and resilience (Seyedin et al., 2024), little research explores its potential to protect healthcare workers from psychosocial risks, particularly in the Moroccan context. This gap motivates our research question: What is the contribution of machine learning to the prevention of emotional exhaustion (burnout) among healthcare workers in Morocco?

We will shed methodological light on our central question based on the Digital HRM theory developed by Bondarouk & Stone (2015) and the work of Mathivet (Frécon & Kazar, 2009), (Mathivet, 2017), (Biernat & Lutz, 2015) and (Heaton, 2015) and (Amini, 2015, p.1-3). The prediction model for emotional exhaustion disorders (burn-out) is created from a conceptual research model inspired by the results of empirical studies by Kapasa et al. (2021) and Fattahi et al. (2023), the transformation of this model into structural equation modeling and the development of a learning algorithm.

This article is organized into four distinct parts. The first part is dedicated to the literature review, allowing us to define the key concepts of our central question and to develop our conceptual research model. The second part presents the research methodology adopted. The third part presents the results of the simulation of the execution of the gradient algorithm. Finally, the fourth part will be devoted to the discussion of our results with regard to previous work.

## 2. Literature review

### 2.1 Apprentissage machine (Machine learning)

Any system created by a healthcare institution, with an adaptive capacity allowing it to react appropriately to its environment, is considered intelligent in the sense of Mathivet (2017, p. 19-23) and is called, according to the author, an artificial intelligence (AI). It therefore has the ability to understand a given health situation, to solve the problems related to it, as well as to select relevant solutions and make the best decisions (Frécon & Kazar, 2009, p. 1). Therefore, an intelligent information system used in the healthcare field requires different information technologies capable of ensuring its adaptability functions, such as big data, machine learning, and others.

In fact, machine learning in healthcare requires a learning algorithm based on modeling a real-life healthcare problem, using one of four main approaches: data classification, regression analysis, clustering, and time series (Biernat & Lutz, 2015, pp. 10-11; Heaton, 2015, p. 7).

In this sense, machine learning in the healthcare sector will allow us to design and develop predictive models capable of automating the execution of tasks that were previously cognitive and performed by humans, without prior programming (Amini, 2015, pp. 1-3). For example, a programmer could code a machine to display a burnout risk message if the workload of healthcare staff exceeds a predetermined weekly hourly volume. In the case of machine learning, through a learning algorithm, the machine would be able to analyze historical data on the workload of healthcare staff and the level of burnout, in order to predict the latter based on the workload. Furthermore, since the characteristics of healthcare staff differ from one individual to another, machine learning can also be used to make personalized decisions regarding the workload of each staff member.

## 2.2 Burnout

### 2.2.1 Definition

In this work, we consider that the most relevant approach is to consider human resources as actors with their own thoughts and motivations, a perspective supported by the authors of the human relations school (Eddaou, 2024, p. 12). In this sense, the mobilization of human resources in health establishments for the implementation of the strategy is not without consequences on the mental, physical and relational health of individuals at work. It is therefore essential to take this dimension into account in order to protect healthcare staff against psychosocial risks.

In this work, we understand psychosocial risks in healthcare facilities to be the various tensions—such as external and internal violence, moral harassment, discrimination, etc.—generated by the implementation of the strategy (Haubold, 2010, pp. 15-24; Clerc, 2018, p. 28). According to these two authors, a psychosocial risk corresponds to an unwanted event that a hazard is likely to cause and that would have negative effects on a target (Le Ray, 2015, pp. 6-7). For example, if a hazard is present in a healthcare facility—such as violent patients requiring urgent intervention—and the attending physician is late, an unwanted event (UE) could occur, such as verbal or physical aggression against nurses (external violence), with consequences for the physical, mental, and relational health of the healthcare staff.

What could happen if these different tensions reach the target (care staff) are physical consequences (sleep problems, weight disorders, etc.), psychological consequences (depressed mood, hopelessness) and behavioral consequences (absenteeism, drug addiction, etc.), leading to a worsening of psychosocial disorders (chronic stress, professional exhaustion "burn-out", etc.) (Haubold, 2010, p.14-15; Clerc, 2018, p.28).

Emotional exhaustion (burnout), as a psychosocial disorder, includes an affective dimension linked to emotional exhaustion, a physical dimension linked to depersonalization, and a psychological dimension linked to the reduction of personal accomplishment (Faye-Dumanget et al., 2018, November, p.2; Kapasa et al., 2021, p.530).

### 2.2.2 The determinants of Burn-Out

**Figure 1: The conceptual research model of healthcare staff burnout**

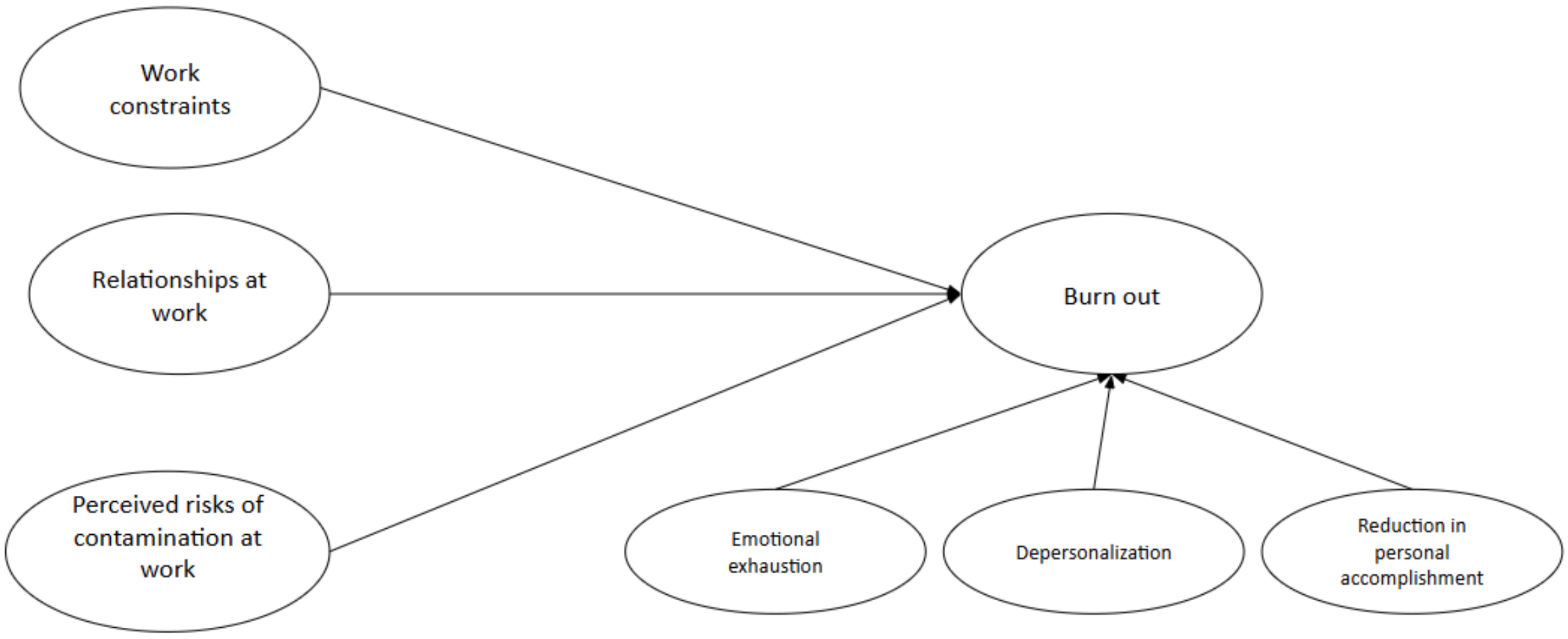

**Source: Established by us**[1]

According to Kapasa et al. (2021, p.526), the statistically significant causes of burnout among healthcare staff during the COVID-19 period are work-related constraints (number of shifts per month, number of working hours per week, number of working hours per day, time occupied), relationships at work (satisfaction with professional relationships with line managers) and perceived risks of contamination at work.

## 2.3 Machine learning and burnout: Theoretical foundations

Based on the Digital HRM theory developed by Bondarouk and Stone (2015), the digitalization of the HR function would improve decision-making regarding working conditions, thus promoting the protection of healthcare staff against the risks of burnout.

# 3. Methodology

## 3.1. Methodological approach

This work is part of a modeling approach and has as its research object a methodological framework centered on the creation of a decision-making tool concerning the determining factors of burnout among healthcare personnel. It aims to propose a model for predicting the risks of emotional exhaustion (burnout), the parameters of which will be estimated using supervised learning.

The burnout risk prediction model is developed from a conceptual research model inspired by the results of the empirical studies of Kapasa et al. (2021) and Fattahi et al. (2023), then transformed into structural equation modeling and supplemented by the development of a learning algorithm.

[1] based on the paper by (Kapasa et al., 2021,p.526)

## 3.2.Variables of the conceptual model of research on Burn-out

**Table No. 1: Explanatory variables of burnout**

| Latent variable | Manifest variable | Ladder | Source |
|---|---|---|---|
| Work constraints (CT) | CT1: The number of guards per month during a reference period | 1. < 5 guards<br>2. 5 to 10 guards<br>3. > 10 guards | (Kapasa et al., 2021, p.526) |
| | CT2: The number of hours worked per week during a reference period | 1. < 40 hours<br>2. > 40 hours | |
| | CT3: The number of hours worked per day during a reference period | 1. ≤ 8 hours<br>2.> 8 hours | |
| | CT4: Occupation time | 1. Permanent/full-time<br>2.Replacement/temporary | |
| Relationships at work (RT) | RT1: Satisfaction with the professional relationship with hierarchical superiors | 1. Not satisfied<br>2. Moderately satisfied<br>3. Satisfied | |
| | RT2: Satisfaction with professional relationships (with colleagues) | 1. Not satisfied<br>2. Moderately satisfied<br>3. Satisfied | |
| Perceived risks of contamination at work (RC) | RC1: Fear of being infected with a serious illness | 1. Restless<br>2. Not worried | |

**Source: Prepared by us[2]**

The table below presents the measurement items of the explanatory latent variables of our conceptual research model, as well as the different measurement scales that correspond to them.

[2]based on the work of Kapasa et al. (2021) and

**Table 2: Manifest variables of the latent variable burn-out (BO)**

| Latent variable | Variable manifeste (Item) | Ladder | Source |
|---|---|---|---|
| Emotional exhaustion (EE) | EE1: I feel emotionally exhausted from work | Liked a 7 points | (Brady et al., 2020, p.3-5) |
| | EE2: I feel exhausted at the end of the workday. | Liked a 7 points | |
| | EE3: I feel tired when I get up and face a new day at work. | Liked a 7 points | |
| | EE4: I feel exhausted from work | Liked a 7 points | |
| | EE5: Working with people all day is a real constraint | Liked a 7 points | |
| | EE6: I feel frustrated with my work | Liked a 7 points | |
| | EE7: I feel like I work too hard. | Liked a 7 points | |
| | EE8: Working directly with people puts too much stress on me | Liked a 7 points | |
| | EE9: I feel exhausted (Empty of energy to continue) | Liked a 7 points | |
| Depersonalization (DP) | DP1: You treat patients as impersonal objects. | Liked a 7 points | |
| | DP2: I have become more insensitive to people | Liked a 7 points | |
| | DP3: I feel emotionally hardened | Liked a 7 points | |
| | DP4: I don't care what happens to some patients. | Liked a 7 points | |
| | DP5: Patients blame you | Liked a 7 points | |
| Reduction in personal accomplishment (PA) | PA1: I easily understand what my patients feel. | Liked a 7 points | |
| | PA2: I deal with patients' problems very effectively. | Liked a 7 points | |
| | PA3: I positively influence the lives of others through my work. | Liked a 7 points | |
| | PA4: I feel full of energy. | Liked a 7 points | |
| | PA5: I can easily create a relaxed atmosphere with patients | Liked a 7 points | |
| | PA6: I feel the absence of a sense of elation after working closely with patients. | Liked a 7 points | |
| | PA7: I accomplish many interesting things at work. | Liked a 7 points | |
| | PA8: I handle emotional issues related to work very calmly. | Liked a 7 points | |

**Source: Prepared by us[3]**

The table below illustrates the measurement items of the latent variables to be explained in our conceptual research model, as well as the related 7-point Likert scale.(0 = never, 1 = a few times a year or less, 2 = once a month or less, 3 = a few times a month, 4 = once a week, 5 = several times a week, 6 = every day).

**3.3. Modeling of structural equations on Burn-out**

**→ Structural equation modeling**

[3]based on the work of Kapasa et al. (2021).

**Figure 2: Structural equation modeling**

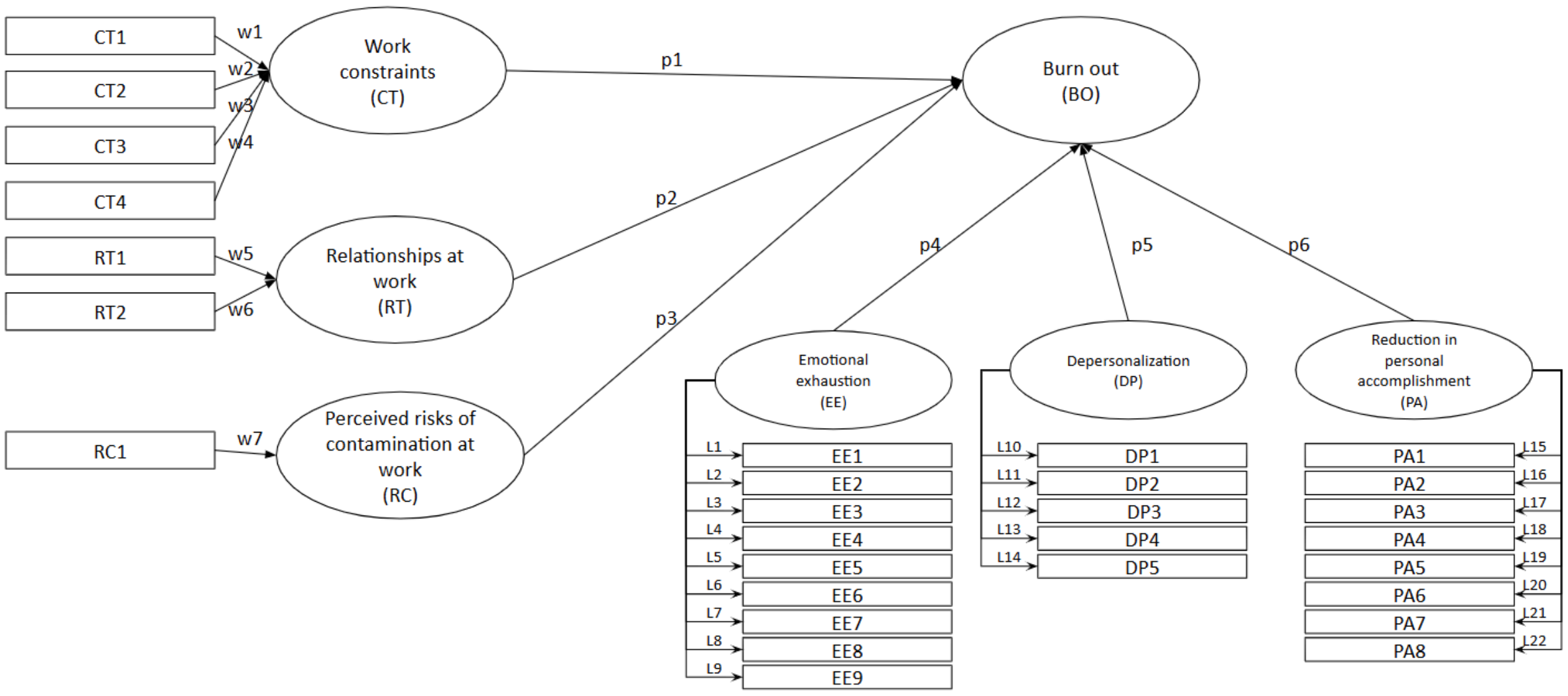

**Source: Prepared by us[4]**

To successfully perform the empirical analysis of our structural equation model, we will use a two-step analysis as explained by (Hair et al., 2017).

**Figure 3: Structural equation modeling**

**Step 1: Specification of the hierarchical model and estimation of the factor scores of the lower-level components**

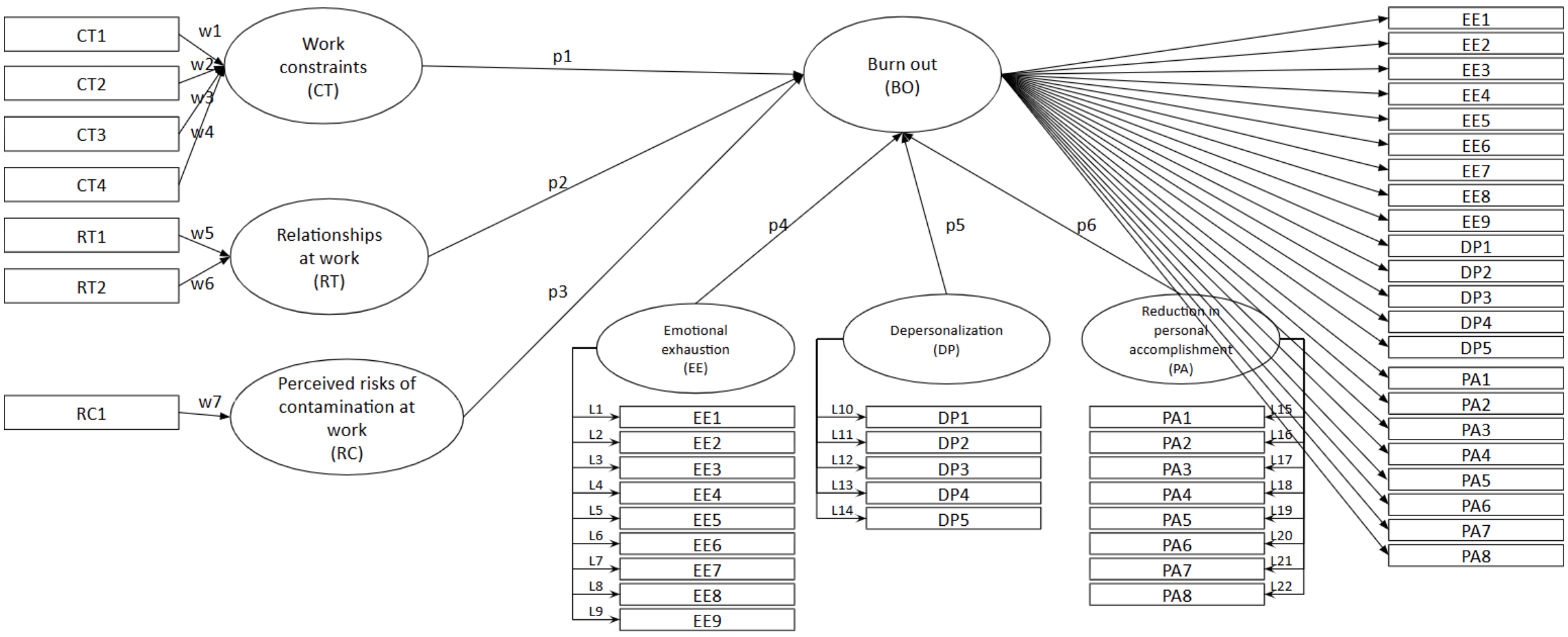

## Step 2: Estimation of the parameters of the structural equation model

[4] based on our literature review, the table of explanatory latent variables and the latent variable to be explained and based on the work of Hair et al. (2017).

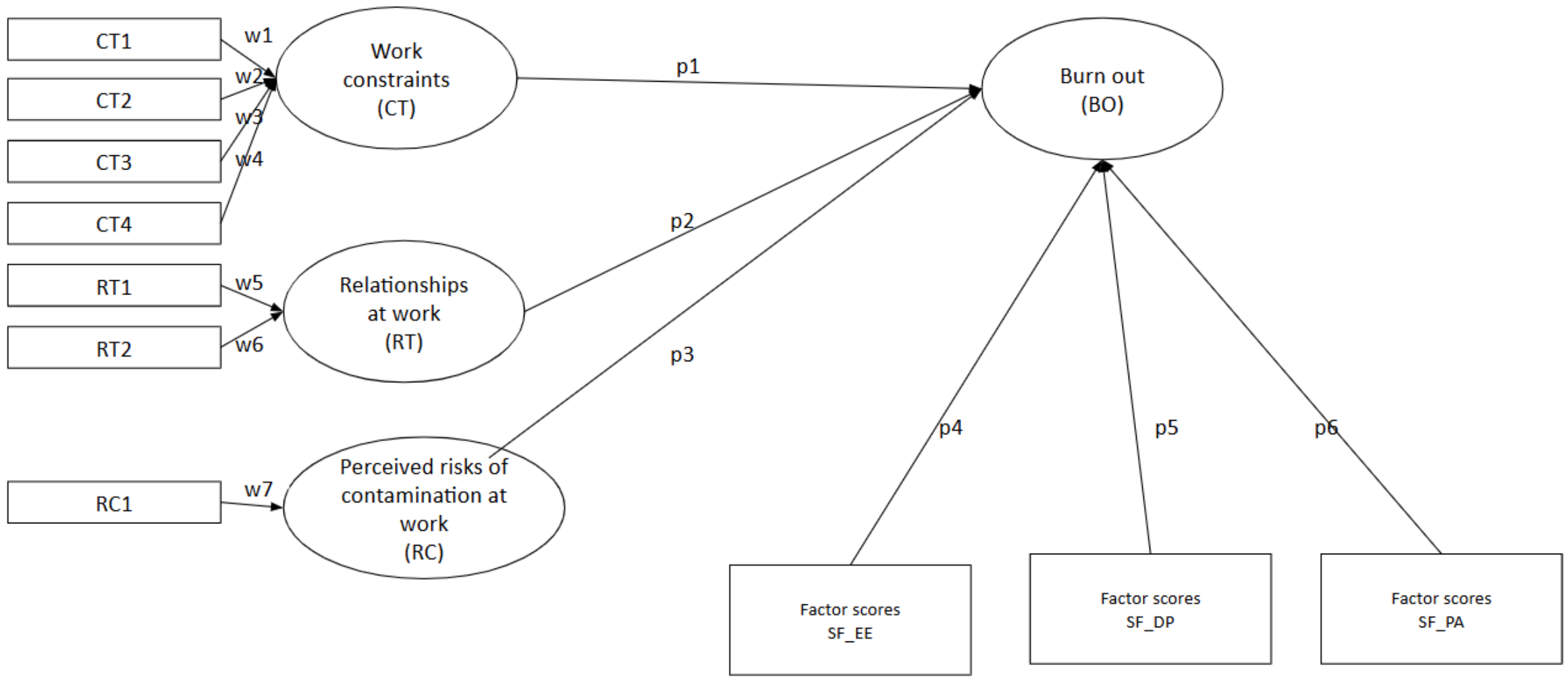

**Source: Prepared by us[5]**

This analysis is suitable for hierarchical component models (HCM). As illustrated in the figure above, the empirical analysis of a hierarchical component model is carried out in two steps. The first step is to associate the measurement indicators of the lower-level components (EE, DP, PA) with the higher-level component (BO), and then to estimate the factor scores of the lower-level constructs (EE, DP, PA) so that they become manifest variables of the higher-level construct (BO). The second step is to estimate the model resulting from the first step.

Approach:

- Step 1: Estimate the MCH parameters by covariance-based structural equation modeling (CB-SEM), using SPSS AMOS software.
- Step 2: Calculate the factor scores of the lower-level components (EE, DP, PA) with the imputation regression method of SPSS AMOS, and then name them SF_EE, SF_DP, and SF_PA, respectively.
- Step 3: Using simplified structural equation modeling, estimate the parameters of the learning models via linear regression with gradient descent.

In the case where the values of the manifest variables do not follow a normal distribution, it is preferable to estimate the parameters of the MCH by using structural equation modeling based on partial least squares SEM-PLS and using SMART PLS software.

### 3.4. Supervised Learning: Approach

In our research, machine learning in a healthcare setting involves providing the machine with input (data on factors explaining burnout) and output (data on burnout sub-scores)

[5]based on the work of Hair et al. (2017).

pairs so that it can learn a function linking the factors to the burnout sub-scores (Russell & Norvig, 2010, p.737). In this case, it is a regression problem.

To successfully perform supervised learning and solve our regression problem, we will, in this work, use the gradient algorithm (Amini, 2015, p.41). This approach will allow us to estimate the parameters of our burnout prediction model.

Based on the work of Raykov and Marcoulides (2006, p.15) and our structural equation modeling, the assumptions of the learning models in this work are as follows:

Structural model no. 1: $BO_i = p1.CT_i + p2.RT_i + p3.RT_i + e1_i$

Structural model no. 2: $BO_i = p4.SF_EE_i + p5.SF_DP_i + p6.SF_PA_i + e2_i$

Formative measurement model no. 1: $CT_i = w1.CT1_i + w2.CT2_i + w3.CT3_i + w4.CT4_i + e3_i$

Formative measurement model no. 2: $RT_i = w5.RT1_i + w6.RT2_i + e4_i$

Formative measurement model no. 3: $RC_i = w7.RT1_i + e5_i$

With :

- i: the number of the respondent to the questionnaire.

With :

- $p_i$: the weights of the structural model,
- $In_i$: the weights of the formative measurement models,
- $L_i$: the factor loadings of reflective measurement models,
- $and_i$ is the random error of the structural model and the measurement models.

Based on Brown (2015, p.62), the cost function corresponds to the following maximum likelihood fitting function:

$$F_{ML} = ln|S| - ln|\Sigma| + trace[(S)(\Sigma^{-1})] - p$$

With :

- $|S|$ is the determinant of the input variance-covariance matrix,
- $|\Sigma|$ is the determinant of the predicted variance-covariance matrix
- p is the order of the input matrix.

The input variance-covariance matrix $S$ :

| | CT1 | CT2 | CT3 | CT4 | RT1 | RT2 | RC1 | SF_EE | SF_DP | SF_PA |
|---|---|---|---|---|---|---|---|---|---|---|
| CT1 | V1 | | | | | | | | | |
| CT2 | CV21 | V2 | | | | | | | | |
| CT3 | CV31 | CV32 | V3 | | | | | | | |
| CT4 | CV41 | CV42 | CV43 | V4 | | | | | | |
| RT1 | CV51 | CV52 | CV53 | CV54 | V5 | | | | | |
| RT2 | CV61 | CV62 | CV63 | CV64 | CV65 | V6 | | | | |
| RC1 | CV71 | CV72 | CV73 | CV74 | CV75 | CV76 | V7 | | | |
| SF_EE | CV81 | CV82 | CV83 | CV84 | CV85 | CV86 | CV87 | V8 | | |
| SF_DP | CV91 | CV92 | CV93 | CV94 | CV95 | CV96 | CV97 | CV98 | V9 | |
| SF_PA | CV101 | CV102 | CV103 | CV104 | CV105 | CV106 | CV107 | CV108 | CV109 | V10 |

The predicted variance-covariance matrix$\Sigma$ :

$$\Sigma = \Lambda.(I - B)^{-1}.\Psi.(I - B)^{-T}.\Lambda^{-T} + \Theta$$

With :

- $\Lambda$: Internal weight matrix$In_i$with i ranging from 1 to 7.
- $B$: External weight matrix$p_i$with i ranging from 1 to 6.
- $\Psi$: Variance-covariance matrix of latent variables
- $\Theta$: Variance-covariance matrix of measurement errors

The objective is to estimate the parameters of our learning model on burnout, in order to produce a predicted variance-covariance matrix (denoted Σ) which is as close as possible to the observed variance-covariance matrix (denoted S).

Based on the work of Amini (2015, p.41), the update of the parameters$v_i$is carried out by using the following gradient descents:

$$\forall t \in N,\ v_i^{(t+1)} = v_i^{(t)} - \eta.\frac{\partial F_{ML}}{\partial v_i^{(t)}}$$

With :

- $\eta$the learning rate,
- $\frac{\partial F_{ML}}{\partial v_i^{(t)}}$are the gradients of the cost function with respect to the parameter$v_i$.
- $v_i^{(t)}$is the value of parameter i at iteration t,

- $v_i^{(t+1)}$ is the updated value of parameter i at the next iteration.

List of parameters:

- Weight of structural models:p1, p2, p3, p4, p5, p6.
- Weights of formative measurement models: w1, w2, w3, w4, w5, w6, w7.

The partial derivative of $\frac{\partial F_{ML}}{\partial v_i^{(t)}}$ :

$$\frac{\partial F_{ML}}{\partial v_i^{(t)}} = - trace\left[(\Sigma^{-1} + \Sigma^{-1}.S.\Sigma^{-1}).\frac{\partial \Sigma}{\partial v_i^{(t)}}\right]$$

And $v_i^{(t)}$ = w1 (weight of CT1 on CT)

1.Build $\frac{\partial \Lambda}{\partial w1}$ (sparse matrix with a 1 in position CT1 → CT)

2. Calculate $\frac{\partial \Sigma}{\partial w1}$

3.insert into $\frac{\partial F_{ML}}{\partial w1}$

4. Calculate $\frac{\partial F_{ML}}{\partial w1}$

5. Update $In1^{(t+1)} = w1^{(t)} - \eta.\frac{\partial F_{ML}}{\partial w1}$

Steps for running the gradient algorithm:

- Start by giving initial values to the parameters and η at iteration t
- Calculate the value of $F_{ML}$
- If the value of of $F_{ML}$ is not minimal, this means that the gap between the predicted variance-covariance matrix (denoted Σ) and the observed variance-covariance matrix (denoted S) is still significant.
- Update settings $v_i^{(t+1)}$ by applying gradient descent.
- Repeat this process until the value of the cost function $F_{ML}$ be minimized. Additional iterations will allow this minimum value to be reached.

To simplify the simulation, we used the following approach:

1. Start by giving initial values to the parameters $v_i$ And η at iteration t =0

2. Calculate the value of $F_{ML}$

3. Update settings $v_i^{(1)}$ by applying gradient descent.

4. Repeat until iteration t = 1

5. Add a prediction of SF_EE, SF_DP and SF_PA item scores based on the input values.

- CT1: The number of guards per month during a reference period
    - 3. > 10 guards
- CT2: The number of working hours per week during a reference period
    - 2. > 40 hours
- CT3: The number of hours worked per day during a reference period
    - 2.> 8 hours
- CT4: Occupation time
    - 1. Permanent/full-time
- RT1: Satisfaction with the professional relationship with hierarchical superiors
    - 1. Not satisfied
- RT2: Satisfaction with professional relationships (with colleagues)
    - 1. Not satisfied
- RC1: Fear of being infected with a serious illness
    - 1. Restless (e)

6. Transform the scores obtained to fit a 7-point Likert scale (0 = never, 1 = a few times a year or less, 2 = once a month or less, 3 = a few times a month, 4 = once a week, 5 = several times a week, 6 = every day). We used the following formula for transforming the EE scores:

$$a = \frac{7}{max(EE) - min(EE)}$$

$$b = 1 - a * min(EE)$$

$$EE_{7points} = a * EE + b$$

7.Prsubmit a comment respecting the following rule:

- A higher burden of exhaustion symptoms is manifested by high scores on the Emotional Exhaustion (EE) and Depersonalization (DE) subscales.
- Conversely, lower scores on the personal accomplishment (PA) subscale also indicate increased exhaustion symptom burden.

**3.5. Presentation of the simulation of supervised learning**

After running the gradient algorithm and estimating the parameters of our learning models - without respecting the expected values of the descriptive indices of goodness of fit, in particular the Chi-square ($\chi^2$) test (Brown, 2015, p. 67) - we begin the phase of predicting burnout scores. This step aims to identify healthcare employees at high risk of burnout.

Score prediction approach:

- Collect scores for manifest explanatory variables.
- Calculate the scores of the latent explanatory variables.
- Calculate the scores of the variable to be explained.
- Calculate the scores of the manifest variable to be explained.
- Infer: High scores on the Emotional Exhaustion (EE) and Depersonalization (DE) subscales, as well as low scores on the Personal Accomplishment (PA) subscale, all indicate a higher burden of exhaustion symptoms.

To demonstrate the practical utility of our burnout predictive model and its potential application in managerial decision-making, we conducted a simulation based on fictitious data. This case study involves a simulated sample of 2,000 Moroccan healthcare professionals, allowing:

- Estimating the parameters of our different learning models,
- Calculation of the predictive values of the latent variables composing the construct of burnout (BO).

We estimated the parameters of our learning models and predicted the scores of the manifest variables of our explanatory variable BO using R code running on the R application. To facilitate the execution of the algorithm, we chose a number of iterations t=1

## 4.Result

### 4.1 Estimation of the parameters of the learning models

**Step 3: Estimation of the parameters of the learning model**

```
--- Itération 1 ---
Paramètres Estimés:
  p (BO sur CT, RT, RC, SF_EE, SF_DP, SF_PA): 0.0646, 0.1598, 0.0877, 0.1778, 0.1887, 0.0187
  w_ct (CT1, CT2, CT3, CT4): 1.0000, 0.1103, 0.1796, 0.1148
  w_rt (RT1, RT2): 1.0000, 0.0968
  w_rc (RC1): 1.0000
  Variances Latentes (CT, RT, RC): 0.4806, 0.2540, 0.3549
  Variances Observées (SF_EE, SF_DP, SF_PA): 0.3077, 0.0964, 0.4549
  Variances Résiduelles Indicateurs (CT1-RC1, partiel): 0.0689, 0.1976, 0.4795, 0.4503, 0.3618 ...
  Variances Résiduelles BO : 0.1607
  Coût (FML) : 23.0800
  Norme du gradient : 1.001610e+02

Modèles d'apprentissage (Formulation - Figure n°2):
  Modèle structurel BO: BO = 0.0646*CT + 0.1598*RT + 0.0877*RC + 0.1778*SF_EE + 0.1887*SF_DP + 0.0187*SF_PA + e_BO
  Modèle de mesure CT: CT1 = 1.0000*CT + e_CT1, CT2 = 0.1103*CT + e_CT2, CT3 = 0.1796*CT + e_CT3, CT4 = 0.1148*CT + e_CT4
  Modèle de mesure RT: RT1 = 1.0000*RT + e_RT1, RT2 = 0.0968*RT + e_RT2
  Modèle de mesure RC: RC1 = 1.0000*RC + e_RC1
Descente de Gradient SEM terminée.
```

**Source: Prepared by us[6]**

[6]using R software

Running the R script implementing the gradient descent algorithm produced the results shown above.

## 2 Prediction of Burnout Scores

**Step 4: Estimation of the learning model parameters**

--- Entry requirements for the hypothetical individual —

```
CT1 CT2 CT3 CT4 RT1 RT2 RC1
  3   2   2   1   1   1   1
```

--- Standardized Hypothetical Inputs ---

```
       CT1        CT2        CT3        CT4        RT1        RT2        RC1
 1.1976116  0.9828963  0.9663172 -0.9653502 -1.2153768 -2.9715069 -1.0097980
```

**Source: Prepared by us**[7]

Running the R code, which implements the gradient algorithm on the R application, generated the results shown in the figure above, as well as the following:

--- Entry requirements for the hypothetical individual —

```
CT1 CT2 CT3 CT4 RT1 RT2 RC1
  3   2   2   1   1   1   1
```

--- Standardized Hypothetical Inputs ---

```
       CT1        CT2        CT3        CT4        RT1        RT2        RC1
 1.1976116  0.9828963  0.9663172 -0.9653502 -1.2153768 -2.9715069 -1.0097980
```

--- Inferred First-Order Latent Factor Scores (Approximation) —

- Latent Factor CT (Work Constraints): 0.5454
- Latent Factor RT (Work Relations): -2.0934
- Latent Factor RC (Perceived Contamination Risks): -1.0098

--- Prediction of Burnout Score (BO) for the hypothetical individual ---

-BO predicted (standardized) = -0.3878

--- Transformation Coefficients of Scores into 7-point Scale (0-6) ---

- SF_EE : a = 1.0000, b = 0.0000
- SF_DP : a = 1.0000, b = 0.0000
- SF_PA : a = 1.0000, b = 0.0000

Average score:

- Observed mean score of SF_EE (original scale): 2.9690

[7]using R software

- Observed mean score of SF_DP (original scale): 2.9450

Transformation of an observed average score for illustration:

- Transformed score of SF_DP (scale 0-6): 2.9450
- Observed mean score of SF_PA (original scale): 3.0430
- Transformed SF_PA score (scale 0-6): 3.0430

According to these scores:

- For the EE and DP variables: Average scores on the Emotional Exhaustion (EE) and Depersonalization (DP) scales reveal a moderate level of these symptoms, suggesting an intermediate risk of burnout in the subject.
- For the PA variable: An average score on the Personal Accomplishment (PA) scale indicates moderate job satisfaction, reflecting partial protection against burnout.

## 5.Discussion

### 5.1 Contributions

In this work, the development of a predictive model of burnout, based on the design of a conceptual research model and the execution of the gradient algorithm to estimate the parameters of our learning models, represents a theoretical and methodological contribution. This contribution stands out from the work of Clerc (2018) and Kapasa et al. (2021) on burnout, as well as from the studies conducted by Seyedin et al. (2024) regarding the role of AI in protecting individuals at work.

### 5.2 Managerial suggestions

This research suggests that decision-making within Moroccan healthcare institutions, particularly regarding working conditions and professional relations, should be based on simulations carried out using this predictive burnout model.

In this perspective, the development of this model requires a full-fledged function that ensures the centralization of data on healthcare staff, the execution of the gradient algorithm to estimate the parameters of the learning models, and the prediction of the Burn-out Score (BO) for the hypothetical individual. This function also requires the implementation of an intelligent information system, capable of making decisions in real time and based on dynamic data. Thus, healthcare institutions could consider governance of this system to ensure its strategic alignment.

**Conclusion**

During the COVID-19 pandemic, Moroccan public hospitals have experienced a sharp increase in demand, resulting in a significant workload for healthcare workers and raising critical questions about their occupational health.

Despite the interest in artificial intelligence (AI) in military personnel safety, there is a lack of research regarding its application to the protection of healthcare personnel from psychosocial risks, such as burnout.

This gap motivated our study, whose central question is: What is the contribution of machine learning to the prevention of emotional exhaustion (burn-out) among staff of healthcare establishments in Morocco?

The first contribution of this work, compared to previous studies, is theoretical. It lies in the design of a conceptual model of research on burnout, which enriches knowledge of the explanatory factors of psychosocial risks.

The second contribution is methodological. It consists of the mobilization of an empirical analysis tool for the conceptual research model, dedicated to the multivariate analysis of latent variables. This tool includes structural equation modeling, hierarchical component modeling, and the subsequent development of a predictive model of burnout through supervised learning.

This research suggests that decision-making in healthcare institutions in Morocco regarding working conditions and professional relationships should be based on simulations carried out using this predictive burnout model.

As a limitation of this research, it is important to note that our predictive model of burnout is based on a conceptual model that has not yet been empirically validated on a large sample size and associated with goodness-of-fit indices. In addition, the execution of the gradient algorithm requires that the values of the manifest variables follow a normal distribution. This is due to the fact that this algorithm uses a maximum likelihood-based fitting function as a cost function.

The prospects for future research consist of carrying out an empirical study on a significant sample of healthcare personnel in Morocco in order to verify the hypotheses of our conceptual model (EDDAOU, 2025, February 18; EDDAOU, 2025, February 25).